\documentclass{article}
\usepackage{amsmath}
\def\t{\theta}

\def\be{\begin{equation}}
\def\ee{\end{equation}}
\def\arr{\begin{array}{rll}}
\def\ea{\end{array}}
\def\bea{\begin{eqnarray}}
\def\eea{\end{eqnarray}}

\def\N2{$N{=}2$}

\def\>{\rangle}
\def\<{\langle}
\def\+{\dagger}
\def\={\ =\ }

\begin{document}
\renewcommand{\thefootnote}{\fnsymbol{footnote}}
\begin{titlepage}
\setcounter{page}{0}
\begin{flushright}
LMP-TPU--9/11  \\
\end{flushright}
\vskip 1cm
\begin{center}
{\LARGE\bf Conformal mechanics inspired by }\\
\vskip 0.5cm
{\LARGE\bf extremal black holes in $d=4$ }\\
\vskip 2cm
$
\textrm{\Large Anton Galajinsky$\;^{a}$ and Armen Nersessian$\;^{b}$\ }
$
\vskip 0.7cm
$\;^a${\it
Laboratory of Mathematical Physics, Tomsk Polytechnic University, \\
634050 Tomsk, Lenin Ave. 30, Russian Federation} \\
$\;^b${\it Yerevan State University,
1 Alex Manoogian St., Yerevan, 0025, Armenia}\\
{Emails: galajin@mph.phtd.tpu.ru, arnerses@ysu.am}

\end{center}
\vskip 1cm
\begin{abstract} \noindent
A canonical transformation which relates the model of a massive relativistic
particle moving near the horizon of an extremal black hole in four dimensions and the conventional conformal mechanics
is constructed in two different ways.
The first approach makes use of the action--angle variables in the angular sector. The second scheme relies upon
integrability of the system in the sense of Liouville.
\end{abstract}

\vspace{0.5cm}

\end{titlepage}

\renewcommand{\thefootnote}{\arabic{footnote}}
\setcounter{footnote}0

\section{Introduction}

In four dimensions, a stationary, rotating, asymptotically flat vacuum black hole with a single degenerate horizon is essentially unique and is
given by the extremal Kerr solution \cite{ahmr}. The analog of the uniqueness theorem for charged black holes yields the extremal
Kerr--Newman solution \cite{ahmr}. In the near horizon limit the solutions exhibit conformal invariance \cite{bh}
which points at a possible dual description in terms of conformal field theory \cite{ghss}. The Kerr/CFT correspondence provides an important pattern where a dual description of a black hole goes beyond string theory (for a review and further references see a recent work \cite{bkls}).

In general, some important features of a black hole are adequately captured in the model of a relativistic particle moving on the curved
background.
A classic example of such a kind is the discovery of a quadratic first integral for a massive particle in the Kerr space--time \cite{c} which preceded the construction of the second rank Killing tensor for the Kerr geometry \cite{wp}. In some instances the argument can be reversed.
In Ref. \cite{cdkktp} a massive charged particle moving near the horizon of the extremal Reissner-Nordstr\"om black hole was related to the
conventional conformal mechanics in one dimension \cite{aff} by implementing a specific limit in which
the black hole mass $M$ is large, the difference between the particle mass and the absolute value of its charge
$(m-|e|)$ tends to zero with $M^2 (m-|e|)$ fixed. The angular variables effectively decouple in the
aforementioned limit and show
up only in an indirect way via the effective coupling constant characterizing the conformal mechanics.
 In that setting the absence of a normalizable ground state in the conformal mechanics
and the necessity to redefine its Hamiltonian \cite{aff} were given a new black hole
interpretation \cite{cdkktp}.

In the near horizon limit the isometry group of the extremal black holes in four dimensions contains
 $SO(2,1)$ factor \cite{kl}. Because the latter is the conformal group in one dimension,
 the model of a relativistic particle propagating on such backgrounds is automatically conformal invariant.
In principle, by applying a proper coordinate transformation its Hamiltonian can be put
in the conventional conformal mechanics form
\be\label{dec}
H=\frac{p^2_X}{2}+\frac{2 {\cal  I }(u)}{X^2},\qquad
\Omega=dp_X\wedge dX+\frac 12\omega_{\alpha \beta}(u)~du^\alpha\wedge du^\beta.
\nonumber\\
\ee
Here $X$ and $p_X$ are the effective radial coordinate and the momentum,
and the dynamics of the angular variables $u^\alpha$ is governed by the effective coupling ${\cal I}(u)$ which is given by
the Casimir element of $so(2,1)$. The former and
the latter descriptions are designated in the literature as the $AdS$ and conformal bases \cite{ikn}.

A salient feature of the conformal basis is that it gives rise to a reduced
Hamiltonian system governed by the symplectic structure $\frac 12 \omega_{\alpha\beta}du^\alpha\wedge du^\beta$
and the Hamiltonian in the angular sector ${\cal I}(u)$ which is worthy of study by itself. A relation between the original conformal mechanics and
its angular sector is far from trivial. For example, constants of motion of the initial conformal mechanics and those of its
angular part are related in a rather intricate way \cite{cuboct,hkln,hlns}. One can also reverse the argument and
construct new integrable conformal mechanical systems starting from known ones. For instance, the spherical part of the rational $A_N$--Calogero
model leads to the superintegrable multi--center Higgs oscillator on an $(N-1)$--dimensional sphere \cite{cuboct}.
Conformal basis also provides a useful means for supersymmetrization of conformal mechanics. Given
${\cal N}=4$ supersymmetric extension of the Hamiltonian system $(\frac 12 \omega_{\alpha\beta}du^\alpha\wedge du^\beta, {\cal I}(u))$,
one can immediately construct its generalization which enjoys the full $D(1,2;\alpha)$ symmetry \cite{hkln}.

In the context of black hole physics, a coordinate transformation providing a link between the $AdS$ and conformal bases
has been constructed only for a massive charged particle moving near the horizon the
extremal Reissner-Nordstr\"om black hole. First the radial motion of the particle was related to $d=1$ conformal mechanics of \cite{aff}
at the off--shell Lagrangian level \cite{ikn}. Then the analysis was extended to include the angular variables within the framework of the
Hamiltonian formalism \cite{bgik,g1}.
The purpose of this work is to construct such a transformation for the near horizon extremal Kerr, Kerr-Newman,
Kerr-Newman--AdS--dS, and rotating dilaton--axion black holes in four dimensions.

In the next section we demonstrate that the construction of a canonical transformation which relates the
$AdS$ and conformal pictures is straightforward provided one resorts to the action--angle variables in the angular sector of the model.
Because in some cases the explicit construction of the action--angle variables may prove to be technically involved,
in Sect. 3 we discuss an alternative approach which relies upon integrability of the system in the sense of
Liouville. In particular, it leads to explicit transformation rules for all the backgrounds mentioned above.
We illustrate both the methods by the examples of the extremal Reissner-Nordstr\"om black hole
and the extremal rotating dilaton--axion black hole in Sect. 4.
Concluding Sect. 5 contains the summary and the discussion of possible further developments.

\section{Canonical transformation via action--angle variables}

Conformal mechanics associated with the near horizon geometry of an extremal black hole in four dimensions is
described by the triple
\bea\label{ham}
&&
H=r \left( \sqrt{{(r p_r)}^2 +  L(\t,p_\t,p_\phi)} -q(p_\phi) \right), \quad D=tH+r p_r,
\nonumber\\[2pt]
&& K=\frac{1}{r} \left( \sqrt{{(r p_r)}^2 +  L(\t,p_\t,p_\phi)}
+q(p_\phi) \right)+t^2 H+2tr p_r, \eea which involves the
Hamiltonian $H$, the generator of dilatations $D$, and the generator
of special conformal transformations $K$. Under the Poisson bracket
they form $so(2,1)$ algebra
\be\label{confalg} \{H,D
\}=H, \quad \{H,K \}=2D, \quad \{D,K \} =K. \ee
In general, the representation
(\ref{ham}) is defined up to the rescaling
\be
H \rightarrow \lambda
H, \qquad K \rightarrow \frac{1}{\lambda} K.
\ee

The function
$L(\t,p_\t,p_\phi)$ entering Eq. (\ref{ham}) has the special
structure \be\label{L} L(\t,p_\t,p_\phi)=a(\t)+b(\t) p_\t^2+c(\t)
{(p_\phi+f(\t))}^2, \ee where the coefficients $a(\t)$, $b(\t)$,
$c(\t)$, $f(\t)$ and $q(p_\phi)$ depend on the details of a
particular black hole under consideration (see \cite{bgik,g1} for
the near horizon extremal Reissner-Nordstr\"om black hole,
\cite{cg} for the rotating dilaton--axion black hole, \cite{g2}
for the extremal Kerr throat solution, and \cite{g3} for the near
horizon extremal Kerr-Newman and Kerr-Newman--AdS--dS black holes).
Note that setting $t=0$ in $D$ and $K$ above one does not spoil the
algebra (\ref{confalg}). In what follows we denote $\tilde
D={D|}_{t=0}$, $\tilde K={K|}_{t=0}$. It is worth mentioning
that the Casimir element of $so(2,1)$ is given in terms of $L$ and
$q$ \be {\cal I}=HK-D^2=L(\t,p_\t,p_\phi)-q(p_\phi)^2. \ee

Along with the conformal generators, $L(\t,p_\t,p_\phi)$ and $p_\phi$ constitute the conserved charges of a particle which, in principle, allow one
to integrate canonical equations of motion. The dynamics of the radial pair $(r,p_r)$ is fixed algebraically from $H$ and $D$. $p_\phi$ is a constant of the motion, while $L$ determines $p_\t$. The remaining equations of motion for $\t$ and $\phi$ can be integrated by quadratures.

Note that the angular sector of the system under consideration
defined by the reduced Hamiltonian ${\cal I}(\theta, p_\theta, p_\phi)$ and the symplectic structure
$\omega=dp_\theta\wedge d\theta+dp_\phi\wedge d\phi$ is an
integrable system. Because canonical variables in the angular sector take finite values\footnote{
The angular variables take finite values by definition. That the momenta are finite follows from the integrals of motion $L$ and $p_\phi$ which fix their dynamics.}, one
can introduce the action--angle variables $(I_a,\Phi^a)$ such that\footnote{Because $L(\t,p_\t,p_\phi)$ is a conserved quantity, after passing to the action--angle variables it may
depend on $I_1$ and $I_2$ only.}
\be
{\cal
I}=L(I_1,I_2)-q(I_2)^2,\qquad \omega=dI_a\wedge
d\Phi^a,\quad \Phi^a\in[0,2\pi), \quad a=1,2.
\label{sph}
\ee
In these terms the conformal generators acquire the form
\be\label{haa} H={r}\left(\sqrt{{(r p_r)}^2
+L(I_{1},I_2) } -q(I_2)\right),\quad {\tilde K}=\frac{1}{r} \left(\sqrt{
{(r p_r)}^2 +L(I_{1},I_2) } +q(I_2)\right), \quad {\tilde D}=r p_r, \ee
while the symplectic structure reads
\be
\Omega=dp_r\wedge dr+dI_a\wedge d\Phi^a,\quad \Phi^a\in[0,2\pi),
\quad a=1,2.
\ee

At this point, the Hamiltonian can be put in the conventional
conformal mechanics form by introducing the new radial coordinates as in \cite{hkln} (for related earlier studies see \cite{bgik,g1})
\be\label{xp}
X=\sqrt{2 \tilde K}, \qquad P_X=-\frac{2 \tilde D}{\sqrt{2 \tilde K}}, \qquad \{X,P_X\}=1 \quad \Rightarrow \quad H=\frac 12 P^{2}_X+\frac{2 {\cal I}}{X^2},
\ee
with ${\cal I}$ from (\ref{sph}). However, with respect to the Poisson bracket the new radial variables $(X, P_X)$ do not commute
with $\Phi^a$.
In order to split them, let us express $(r,p_r)$ via $(X,P_X, I_a)$
\be
 r=\frac{2}{X^2}\left(\sqrt{{(XP_X)}^2/4
+L(I_1,I_2) } +q(I_2)\right),  \quad p_r=-\frac{P_X
X^3}{4\left(\sqrt{{(XP_X)}^2/4 +L(I_1,I_2) } +q(I_2)\right)}. \ee
 Then one can
 substitute these expressions into
the symplectic two--form to get
\be
\Omega=dP_X\wedge dX+dI_a\wedge d{\widetilde\Phi}^a, \ee where \be
{\widetilde\Phi^a}=\Phi^a+\frac12\int d({XP_X
})\frac{\partial\log\left(\sqrt{{(XP_X)}^2/4 +L(I_1,I_2) } +q(I_2)\right)
}{\partial I_a}. \label{tphi}
\ee
As a result, $(X,P_X)$ and $(\tilde \Phi^a,I_a)$ constitute canonical pairs.

Thus, we have demonstrated that the construction of a canonical transformation which
provides a bridge between the model of
a massive relativistic particle moving near the horizon
of an extremal black hole in four dimensions and the conventional conformal mechanics is
facilitated by resorting to the action--angle variables in the angular sector.

\section{Alternative approach}

In the previous section we presented the general scheme how to construct a canonical transformation which links
the $AdS$ and conformal bases.
However, in some instances explicit derivation of the action--angle variables may constitute a non--trivial problem \cite{arnold}.
In particular, for the black hole configurations mentioned in the Introduction they are unknown in explicit form.
In this section we discuss an alternative possibility which leads to explicit transformation rules. It relies upon the fact that
the system at hand is integrable in the sense of Liouville.

Let us introduce new canonical momenta $F_1$ and
$F_2$ which are related to the conserved charges $L$ and $p_\phi$\footnote{For definiteness, we take
$I_1$ to be positive. The case $I_1=-\sqrt{a(\t)+b(\t) p_\t^2+c(\t)
{(p_\phi+f(\t))}^2}$ can be considered likewise.}
\be\label{mom}
F_1=\sqrt{L}=\sqrt{a(\t)+b(\t) p_\t^2+c(\t) {(p_\phi+f(\t))}^2}, \qquad
F_2=p_\phi. \ee Transformation rules to new canonical coordinates
$\Theta^1$ and $\Theta^2$ are then found by explicit integration of
linear inhomogeneous partial differential equations which follow
from the requirement that the full change of variables be canonical. The method
of characteristics (see e.g. \cite{smirnov}) yields\footnote{For the
case at hand one has to assume $b(\t) p_\t>0$. Another option $b(\t)
p_\t<0$ is realized for $F_1=-\sqrt{a(\t)+b(\t) p_\t^2+c(\t)
{(p_\phi+f(\t))}^2}$.}
\bea
&&
\Theta^1=\sqrt{a(\t)+b(\t)
p_\t^2+c(\t) {(p_\phi+f(\t))}^2}\times
\nonumber\\[2pt]
&&
\qquad \times
 \int_{\t_0}^\t \frac{d \tilde\t}{\sqrt{b(\tilde\t)[L(\t,p_\t,p_\phi)-a(\tilde\t)-c(\tilde\t) {(p_\phi+f(\tilde\t))}^2}]},
\nonumber
\eea
\bea\label{cor}
&&
\Theta^2=\phi-\int_{\t_0}^\t \frac{c(\tilde\t) (p_\phi+f(\tilde\t)) d
\tilde\t}{\sqrt{b(\tilde\t)[L(\t,p_\t,p_\phi)-a(\tilde\t)-c(\tilde\t)
{(p_\phi+f(\tilde\t))}^2}]},
\eea
where $\t_0$ is an arbitrary constant\footnote{By construction, there is a natural ambiguity in the definition of $\Theta^1$ and $\Theta^2$ which is
related to the choice of the lower limit $\t_0$
in the definite integrals (\ref{cor}). Note that the primitive integrals in (\ref{cor}) at the lower limit $\t_0$ yield functions of $L$ and $p_\phi$
which, in our approach, are related to
the new momenta $F_1$ and $F_2$.}.
Given a black hole, one first identifies the coefficients $a(\t)$, $b(\t)$,
$c(\t)$, $f(\t)$, $q(p_\phi)$ and then computes the integrals. In general, both lines in
(\ref{cor}) involve elliptic integrals. The inverse transformation
from $(\Theta^1,F_1)$, $(\Theta^2,F_2)$ to $(\t,p_\t)$,
$(\phi,p_\phi)$ is thus given in terms of elliptic functions.
Among the black hole configurations mentioned in the Introduction,
only the Reissner-Nordstr\"om black hole and the rotating dilaton--axion black hole
yield (\ref{cor}) in terms of elementary functions.
These two exceptional cases are considered in more detail in Sect. 4.
Note that, because the new canonical coordinates $\Theta^1$ and $\Theta^2$
do not enter the transformed
Hamiltonian, the explicit form of the transformation rules (\ref{cor}) does not really matter.

In the new canonical coordinates the conformal generators read
\be
H=r \left( \sqrt{{(r p_r)}^2 + F_1^2} -q(F_2) \right), \quad \tilde K=\frac{1}{r} \left( \sqrt{{(r p_r)}^2 +  F_1^2} +q(F_2) \right), \quad
\tilde D=r p_r.
\ee
At this point the passage to the conventional conformal mechanics can be realized as above which yields the effective coupling
\be\label{g}
{\cal I}=F_1^2-q(F_2)^2.
\ee
It is amazing that in the conformal basis all the details of a
particular black hole under consideration are encoded in a single function $q(F_2)$ which enters the effective coupling.
However, in
contrast to the angular variables $\Phi^a$ which parameterize a torus,  $\Phi^a\in[0, 2\pi)$, the domain of $\Theta^a$, in general,
depends on the initial conditions and/or specific properties of a background. Thus, the price paid for the
universality of the presented canonical transformation is its locality.

\section{Examples}

Let us illustrate the above considerations by the examples of the near horizon extremal
Reissner-Nordstr\"om black hole and the near horizon extremal rotating dilaton--axion black hole
for which the analysis can be done in elementary functions.

In the first case the conformal generators which characterize the particle probe read
\cite{g1}
\bea\label{h}
&&
H=\frac{r}{M^2}\left(\sqrt{(mM)^2+{(rp_r)}^2 +p_\t^2+\sin^{-2}\t
p_\phi^2} +e q\right), \qquad \tilde D=r p_r,
\nonumber\\[2pt]
&&
\tilde K=\frac{M^2}{r} \left(\sqrt{{(mM)}^2+{(rp_r)}^2
+p_\t^2+\sin^{-2}\t p_\phi^2} -e q\right),
\eea
where $m$ and $e$ are the mass and the electric
charge of a particle while $M$ and $q$ are the mass and the electric
charge of the black hole. The angular part is given by
a free particle moving on a two-dimensional sphere
\be
{\cal
I}=p^2_\theta+\frac{p^2_\phi}{\sin^2\theta}+(mM)^2-(eq)^2,\quad
\omega = dp_\theta\wedge d\theta+dp_\phi\wedge d\phi.
\ee

In order to construct the action--angle variables, we introduce the
generating function \cite{arnold} \be S({\cal E}, p_\phi
|\theta,\phi)=p_\phi\phi + \int_{{\cal I}={\cal E}={\rm
const}} d\theta\sqrt{{\cal E}-
(mM)^2+(eq)^2-\frac{p^{2}_\phi}{\sin^2\theta}}, \ee which is then
used to construct the desired canonical pairs \bea &&
\Phi^1=-\arcsin{\left(\frac{I_1+I_2}{\sqrt{(I_1+I_2)^2-I_{2}^2}}\cos{\theta}\right)},
\qquad \qquad I_1=\sqrt{{\cal I} -(mM)^2+(eq)^2}-p_\phi,
\nonumber\\[2pt]
&&
\Phi^2=\varphi-\arcsin{\left(\frac{p_\theta}{\sqrt{(I_1+I_2)^2-I_{2}^2}}\right)}
+\Phi^1, \qquad \quad I_2=p_\phi.
\eea
The effective coupling reads
\be
{\cal I}=(I_1+I_2)^2
+(mM)^2-(eq)^2. \label{v3}
\ee
Note that in the previous studies of the model \cite{bgik,g1} the angular sector of the resulting
conformal mechanics was described by a free particle moving on a two--dimensional sphere.
The guiding principle behind the canonical transformation in \cite{bgik}
was to keep the explicit form of the $so(3)$--generators intact.
In our approach no assumption is made on the form of the $so(3)$--generators and the resulting
Hamiltonian in the angular sector reads simpler.
As a further development of this work it would be interesting to construct the action--angle variables for
the case of
the near horizon extremal Kerr and Kerr-Newman black holes.

In order to illustrate the alternative method, let us consider the near horizon extremal rotating
dilaton--axion black hole \cite{cg1}. This solution to the Einstein--Maxwell--dilaton--axion theory
has $SO(2,1) \times U(1)$ isometry group and can be viewed as interpolating between the near horizon extremal
Reissner-Nordstr\"om black hole and the near horizon extremal Kerr black hole  \cite{cg1}.
For a particle probe propagating on such a background the conformal generators read
\cite{cg}
\bea\label{h1}
&&
H=r \left(\sqrt{m^2+{(rp_r)}^2 +p_\t^2+\sin^{-2}\t
{[p_\phi-e \cos\t]}^2   } -p_\phi \right), \qquad \tilde D=r p_r,
\nonumber\\[2pt]
&&
\tilde K=\frac{1}{r} \left(\sqrt{m^2+{(rp_r)}^2 +p_\t^2+\sin^{-2}\t
{[p_\phi-e \cos\t]}^2   } +p_\phi \right).
\eea
Here $m$ and $e$ are the mass and the electric charge of a particle.
Note that this realization of the conformal algebra looks very similar to that corresponding to
the Bertotti--Robinson solution with a magnetic charge \cite{g1}.
The only difference is in the last terms entering the Hamiltonian and the generator of  special conformal transformations.
These terms break $SO(3)$ symmetry intrinsic to the Bertotti--Robinson solution down to
$U(1)$ in accord with the azimuthal symmetry of the dilaton--axion black hole.

Introducing new momenta in accord with (\ref{mom})
\be\label{mom1}
F_1=\sqrt{m^2+p_\t^2+\sin^{-2}\t
{[p_\phi-e \cos\t]}^2   }, \qquad
F_2=p_\phi,
\ee
one then derives from (\ref{cor}) (where for simplicity we choose $\t_0=\pi/2$)
the  coordinates canonically conjugated to $F_1$ and $F_2$
\bea\label{cor1}
&& \Theta^1=-\frac{F_1}{\sqrt{F_1^2-m^2+e^2}}\left(\arcsin{\frac{(F_1^2-m^2+e^2)\cos{\t}-e F_2}{\sqrt{(F_1^2-F_2^2-m^2)(F_1^2-m^2+e^2)+e^2 F_2^2}}} +
\right.
\nonumber\\[2pt]
&&
\qquad \qquad \qquad \qquad \qquad \qquad
\left.
+\arcsin{\frac{e F_2}{\sqrt{(F_1^2-F_2^2-m^2)(F_1^2-m^2+e^2)+e^2 F_2^2}}}
\right),
\nonumber\\[2pt]
&& \Theta^2=\phi-\arctan{\frac{(F_2 p_\t \sin{\t}+[F_2-e\cos{\t}]\sqrt{F_1^2-F_2^2-m^2})\cos{\t}}{e \cos{\t}(F_2-e\cos{\t})-p_\t^2 \sin^2{\t}-p_\t \sin{\t} \sqrt{F_1^2-F_2^2-m^2}}}.
\eea

To the best of our knowledge, the examples considered in this section together with the case of the near horizon extremal Reissner-Nordstr\"om black hole carrying a magnetic charge
exhaust all the possibilities where a canonical transformation can be constructed in terms of elementary functions.

\section{Conclusion}
To summarize, in this work two related methods were proposed to construct a canonical transformation which
links the model of a massive relativistic particle moving near the
horizon of an extremal black hole in four dimensions and the conventional conformal
mechanics. A key ingredient of the first scheme is the use of the action-angle variables in the angular sector of the system.
Particular importance of this formulation is a precise indication regarding the
(non)equivalence of various dynamical systems (cf. \cite{aa}).
The second approach relies upon integrability of the model in the sense of Liouville.
The advantage of the conformal basis is that it allows one to accumulate all specific information about the original model
in a lower-dimensional integrable system governed by the effective coupling ${\cal I}$. Moreover, with such a formulation at hand, one can reduce the problem of constructing an ${\cal N}=4$ superconformal extension
of the original system to that for the angular part \cite{hkln}.

Turning to further developments, it is tempting to investigate which geometries correspond to
other conformal mechanics models known in the literature first by applying the
inverse canonical transformation and then by reconstructing a metric tensor. Reduced integrable systems corresponding to higher dimensional
near horizon extremal black holes are also worthy of study.
\\

{\large Acknowledgements.}
We thank Vahagn Yeghikyan for valuable discussions and Mikhail Vasiliev for useful comments.
A.G. was supported by the Dynasty Foundation, RF Federal Program "Kadry" under
the contract 16.740.11.0469, and RFBR grant 09-02-00078. A.N. was supported by the Volkswagen Foundation grant I/84 496
and the Armenian State Committee of Science under the grants SCS 11-1c258 and SCS-BFBR 11AB-001.


\vspace{0.5cm}


\begin{thebibliography}{nn}
\bibitem{ahmr}
A. Amsel, G. Horowitz, D. Marolf, M. Roberts, Phys. Rev. D {\bf 81} (2010) 024033, arXiv:0906.2367.
\bibitem{bh}
J. Bardeen, G. Horowitz, Phys. Rev. D {\bf 60} (1999) 104030, hep-th/9905099.
\bibitem{ghss}
M. Guica, T. Hartman, W. Song, A. Strominger, Phys. Rev. D {\bf 80} (2009) 124008,
arXiv:0809.4266.
\bibitem{bkls}
I. Bredberg, C. Keeler, V. Lysov, A. Strominger, {\it Cargese lectures on the Kerr/CFT correspondence},
arXiv:1103.2355.
\bibitem{c} B. Carter, Phys. Rev. {\bf 174} (1968) 1559.
\bibitem{wp} M. Walker, R. Penrose, Commun. Math. Phys. {\bf 18} (1970) 265.
\bibitem{cdkktp}
P. Claus, M. Derix, R. Kallosh, J. Kumar, P.K. Townsend, A. Van Proeyen, Phys. Rev. Lett. {\bf 81} (1998) 4553,
hep-th/9804177.
\bibitem{aff}
V. De Alfaro, S. Fubini and G. Furlan, Nuovo Cim. A {\bf 34} (1974) 569.
\bibitem{kl}
H. K. Kunduri, J. Lucietti, J. Math. Phys. {\bf 50} (2009) 082502,
arXiv:0806.2051.
\bibitem{ikn}
E. Ivanov, S. Krivonos, J. Niederle, Nucl. Phys. B {\bf 677} (2004) 485, hep-th/0210196.
\bibitem{cuboct}
T. Hakobyan,  A. Nersessian, V. Yeghikyan,
J. Phys. A {\bf 42} (2009) 205206, arXiv:0808.0430.
\bibitem{hkln}
T. Hakobyan, S. Krivonos, O. Lechtenfeld, A. Nersessian, Phys.\ Lett. A {\bf 374} (2010) 801, arXiv:0908.3290.
\bibitem{hlns}
T. Hakobyan, O. Lechtenfeld, A. Nersessian, A. Saghatelian, J. Phys. A {\bf 44} (2011) 055205, arXiv:1008.2912.
\bibitem{bgik}
S. Bellucci, A. Galajinsky, E. Ivanov, S. Krivonos, Phys. Lett. B {\bf 555} (2003) 99,
hep-th/0212204.
\bibitem{g1}
A. Galajinsky, Phys. Rev. D {\bf 78} (2008) 044014, arXiv:0806.1629.
\bibitem{cg}
G. Cl\'ement, D. Gal'tsov, Nucl. Phys. B {\bf 619} (2001) 741,
hep-th/0105237.
\bibitem{g2}
A. Galajinsky, JHEP {\bf 1011} (2010) 126, arXiv:1009.2341.
\bibitem{g3}
A. Galajinsky, K. Orekhov, Nucl. Phys. B {\bf 850} (2011) 339,
arXiv:1103.1047.
\bibitem{arnold} V.I. Arnold, {\it Mathematical methods in classical mechanics}, Nauka Publ., Moscow,
1973.
\bibitem{smirnov} V.I. Smirnov, {\it A course of higher mathematics}, Vol. 2, Pergamon Press, 1964.
\bibitem{cg1}
G. Cl\'ement, D. Gal'tsov, Phys. Rev. D {\bf 63} (2001) 124011,
gr-qc/0102025.
\bibitem{aa}
O. Lechtenfeld, A. Nersessian, V. Yeghikyan, Phys. Lett. A {\bf 374} (2010) 4647, arXiv:1005.0464.
\end{thebibliography}
\end{document}